# Quantum Phase Transition in the Twisted Three-Leg Spin Tube

Keisuke Ito[1], Shun Yokoo[1], Kiyomi Okamoto[2] and Tôru Sakai[1, 3, 4, 5]

[1]*Graduate School of Material Science, University of Hyogo,*
*3-2-1 Kouto, Kamigori-cho, Ako-gun, Hyogo 678-1297, Japan*
[2]*Faculty of Engineering, Shibaura Institute of Technology,*
*Minuma-ku Saitama 337-8570, Japan*
[3]*Research Center for New Functional Materials, University of Hyogo,*
[4]*Japan Atomic Energy Agency (JAEA), SPring-8,*
[5]*National Institutes for Quantum and Radiological Science and Technology (QST), SPring-8*

**Abstract**

We investigate the spin-1/2 twisted three-leg antiferromagnetic quantum spin tube in the presence of the easy-plane anisotropy, using the numerical diagonalization of finite-size clusters. We show cases of the existence and the absence of magnetization plateau at 1/3 height of saturation magnetization when the anisotropy is tuned. The phenomenological-renormalization group analysis indicates a quantum phase transition between the 1/3 magnetization plateau phase and the plateauless one. The phase diagram is also presented.

## 1. Introduction

The spin system of a tube-type structure has attracted a lot of interest in the field of the molecule-based magnets. Particularly the spin-1/2 three-leg spin tube is the most important because it has strong frustration and quantum fluctuation[1]. One of interesting features is that regular tube has a spin gap, while the spin-1/2 three-leg spin ladder is gapless[2]. Motivated by the recent discovery of the spin tube [(CuCl$_2$tachH)$_3$Cl]Cl$_2$ (tach = *cis,trans*-1,3,5-triamino-cyclohexane)[3], we investigate the spin-1/2 twisted three-leg spin tube[4].

In the classical picture, the magnetization of the magnetic material is continuously increased with increasing external fields. On the other hand, a magnetic-field-induced spin gap is sometimes predicted to appear because of the quantum effects. It should appear as a plateau of the magnetization curve, namely a magnetization plateau or quantization of magnatization[5,6].

The numerical diagonalization study[5] suggested that the spin-1/2 three-leg spin tube has a 1/3 magnetization plateau. However, the magnetization measurement revealed that the compound [(CuCl$_2$tachH)$_3$Cl]Cl$_2$ has no magnetization plateau[3]. The density matrix renormalization group (DMRG) study[4] indicated that the spin-1/2 twisted three-leg spin tube has no plateau for sufficiently large exchange interaction along the chain.

In this paper, as another reason of the gapless structure, we consider the easy-plane anisotropy which stabilizes the 120-degree order. The 1/3 magnetization plateau is expected to vanish for sufficiently large easy-plane anisotropy. Thus we investigate the spin-1/2 twisted three-leg spin tube with the easy-plane anisotropy, using numerical diagonalization based on Lanczös algorithm.

We have successfully observed cases of the existence and the absence of the magnetization plateau at 1/3 height of the saturation magnetization when the anisotropy is tuned. And the phenomenological-renormalization analysis has clarified a quantum phase transition between the magnetization plateau and plateauless phases at this height. The phase diagram is presented in Fig.5.

## 2. Model

We consider the spin-1/2 twisted three-leg spin tube, shown in Fig.1. This is a good theoretical model of the cluster compound [(CuCl$_2$tachH)$_3$Cl]Cl$_2$ [3]. The Hamiltonian is written as

$$H = J_1 \sum_{i=1}^{3} \sum_{j=1}^{L} (S_{i,j}^x S_{i+1,j}^x + S_{i,j}^y S_{i+1,j}^y + \lambda S_{i,j}^z S_{i+1,j}^z)$$

$$+ J_2 \sum_{i=1}^{3} \sum_{j=1}^{L} (S_{i,j}^x S_{i+1,j+1}^x + S_{i,j}^y S_{i+1,j+1}^y + \lambda S_{i,j}^z S_{i+1,j+1}^z)$$

$$+ J_2 \sum_{i=1}^{3} \sum_{j=1}^{L} (S_{i,j}^x S_{i-1,j+1}^x + S_{i,j}^y S_{i-1,j+1}^y + \lambda S_{i,j}^z S_{i-1,j+1}^z)$$

$$- H \sum_{i=1}^{3} \sum_{j=1}^{L} S_{i,j}^z \qquad (1)$$

where $S_{i,j}$ is the spin-1/2 operator, $J_1$ ($J_2$) denotes the intra- (inter-) triangle coupling which is supposed to be antiferromagnetic, and i (j) represents the label of the rung (leg) direction. Here i = 4 (i = 0) is regarded the same as i = 1 (i = 3). The XXZ anisotropy of the interactions is represented by $\lambda$. It is clear that the unit cell is composed of three spins from Fig.1(b). In this paper, we fix $J_1$ to unity and consider the easy-plane anisotropy $\lambda < 1$) only.

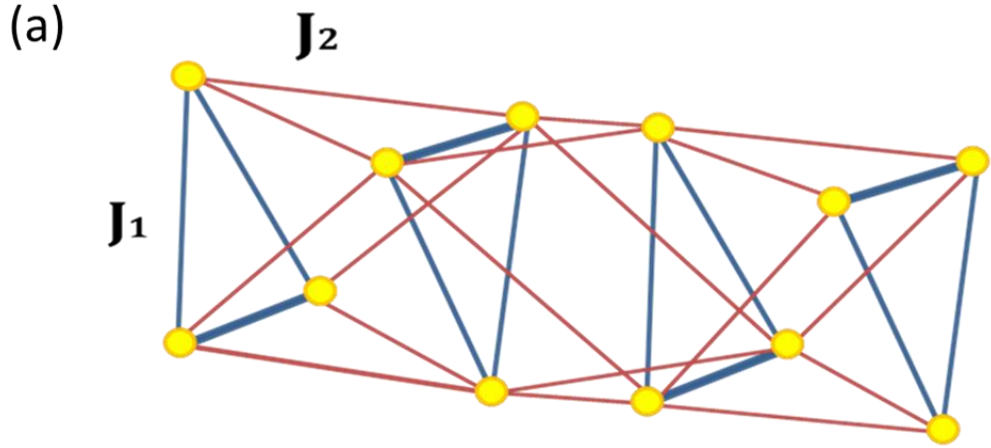

(a)

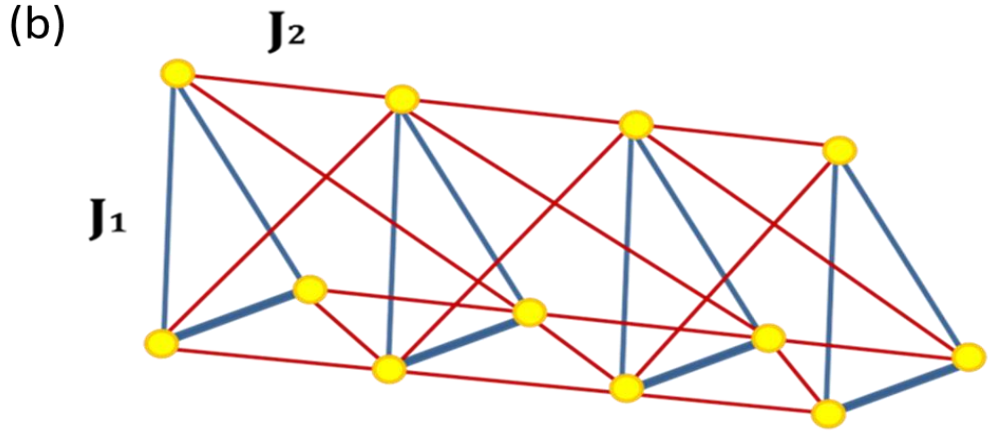

Fig.1 (a) A sketch of the present model in the twisted picture.
(b) Another sketch of the present model in which the space inversion symmetry along the leg direction is clear.

## 3. Magnetization Plateau

Fouet et al. [4] said that the 1/3 magnetization plateau phase is realized for $J_2/J_1 \lesssim 1.5$.

The unit cell consists of three sites as shown in Fig.2. According to Oshikawa-Yamanaka-Affleck theorem, the necessary condition for the existence of magnetization plateau is the form

$$S - m = \text{integer} \qquad (2)$$

where $S$ is the total spin quantum number and $m$ is the magnetization per unit cell. Thus the three-leg spin tube possibly has the 1/3 magnetization plateau even without spontaneous translational symmetry breakdowns. The expected mechanism for the 1/3 plateau is shown in Fig.2(a). On the other hand, the easy-plane anisotropy tends to stabilize the 120° structure like Fig.2(b)[7]. Therefore, a quantum phase transition of the Berezinskii-Kosterlitz-Thouless (BKT) type between the plateau and gapless phases is expected to occur when $\lambda$ decreases[8,9].

In this paper, we fixed $J_1 = 1.0$ and changed $J_2$ (0.2,···,0.5). We performed the numerical diagonalization based on the lanczös algorithm to calculate the lowest energy eigenvalue in each subspace characterized by $M = \sum_j S_j^z$ ($M = 0, ···, 3L/2$), which denoted as $E_0(L, M)$. Using the calculated eigenvalues, the quantum phase transition is investigated.

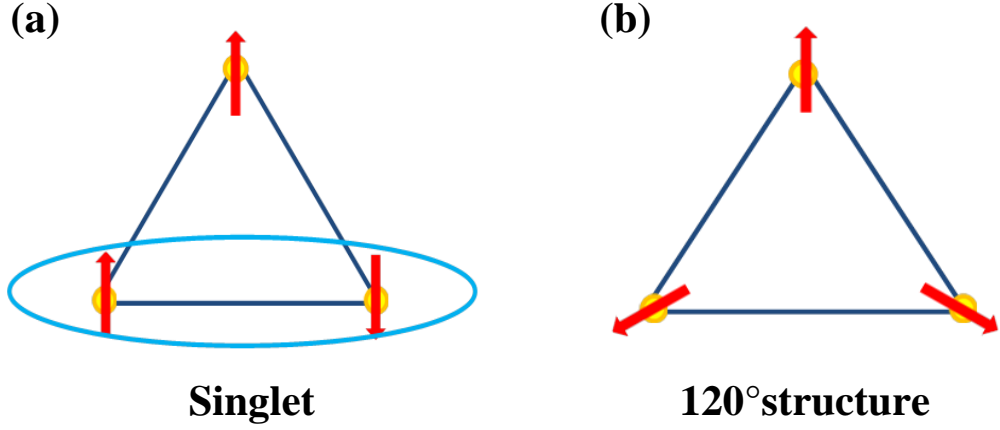

Fig.2 (a) An expected mechanism to realize the 1/3 magnetization plateau. (b) 120-degree structure expected to be realized for a strong easy-plane anisotropy.

## 4. Critical Point

The width of the 1/3 magnetization plateau $\Delta$ is calculated as
$$\Delta = E_0(L, M + 1) + E_0(L, M - 1) - 2E_0(L, M) \tag{3}$$
where $M = L/2$ is 1/3 of the saturation magnetization($3L/2$). Using the numerical diagonalization, we calculate $\Delta$ for $L = 4, 6$ and 8. We numerically confirmed the uniqueness of the lowest energy states in the subspaces of $M = (1/3)M_s$ and $M = (1/3)M_s \pm 1$ except for the $J_2 = 0$ case[10]. Accoriding to the phenomenological renormalization, the equation of the scaled gaps
$$L \Delta (L, \lambda) = (L + 2) \Delta (L + 2, \lambda) \tag{4}$$
gives the size-dependent fixed point $\lambda_c(L + 1)$.

The scaled gap $L\Delta$ is plotted verus $\lambda$ for $L = 4, 6$ and 8. The cross point of the scaled gaps for $L$ and $L+2$ gives the fixed point $\lambda_c(L + 1)$. The scaled gap for $J_2 = 0.2$ is plotted in Fig.3.

The scaled gaps for $L = 6$ and 8 cross to each other and we can obtain $\lambda_c(7)$, but the ones for $L = 4$ and 6 do not cross due to the finite-size effect. Thus we use $\lambda_c(7)$ as the estimated phase boundary between the plateau and plateauless phases at 1/3 of the saturation magnetization. And we performed the same method to obtain $\lambda_c(7)$ for various values of $J_2$.

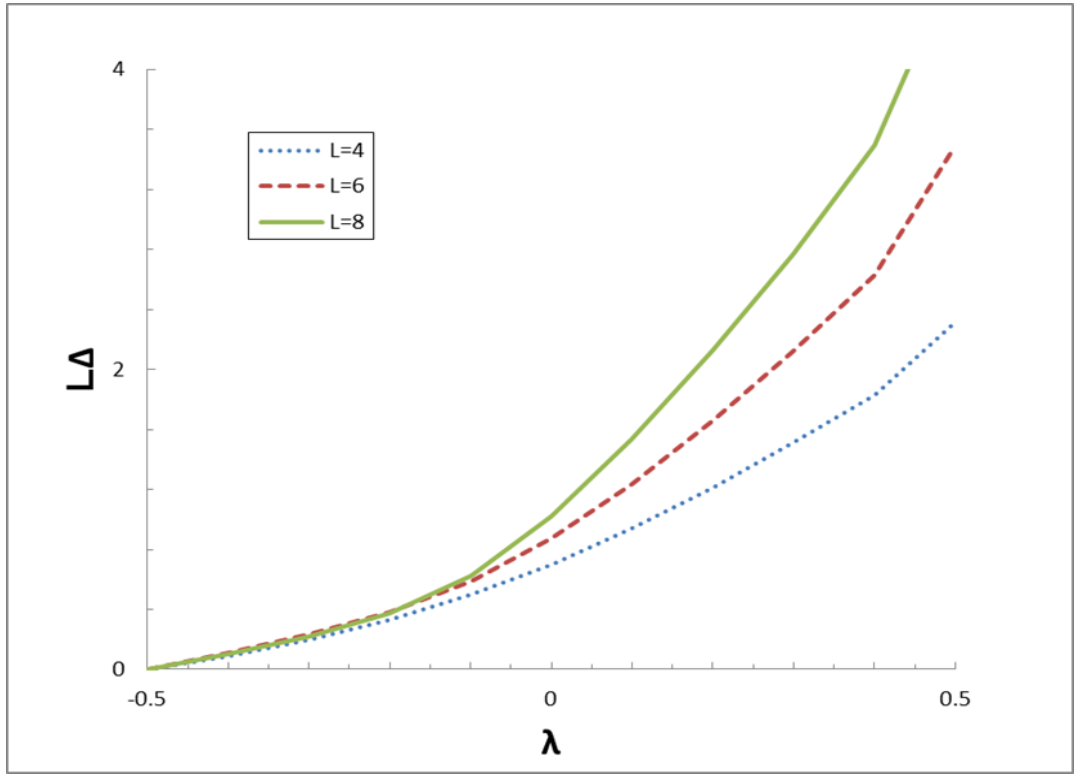

Fig.3 The scaled gap $L\Delta$ is plotted versus $\lambda$ for $L = 4, 6$ and $8$ ($J_2 = 0.2$).

## 5. Phase Diagram

The boundary between the 1/3 plateau and plateauless phases is shown in Fig.4. Also we can consider that the behavior of scaled gaps for $L = 4$ and $6$ indicates the existence of the plateau in the whole region due to a serious finite size effect. Since the phase boundary was obtained only for one system size ($L = 7$), we could not estimate the boundary in the thermodynamic limit. The caluculation for larger systems would be desirable. However, even within the present analysis, we can expect the region of the plateauless phase should be larger in the thermodynamic limit because the finite-size effect overestimates the plateau phase. Therefore we can consider that the present estimation of the phase boundary would give a lower limit of the correct one in the phase diagram.

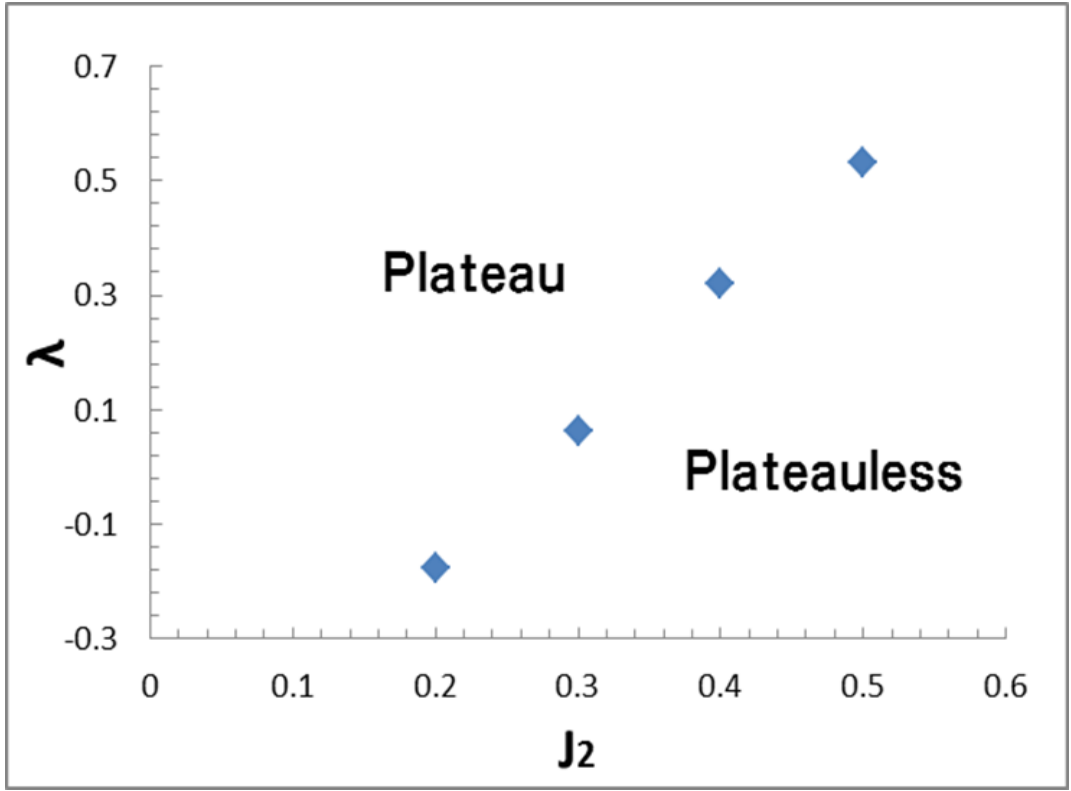

Fig.4 The phase diagram in the $\lambda$-$J_2$ plane. Since the finite size effect tends to overestimate the plateau phase, the present result is expected to give a lower limit of the phase boundary in the thermodynamic limit.

## 6. Discussion and Summary

We have investigated the spin-1/2 twisted three-leg antiferromagnetic quantum spin tube in the presence of the easy-plane anisotropy, using the numerical diagonalization of finite-size clusters and the PRG analysis. It has been pointed out that the PRG analysis for the BKT transition often brings about the overestimation of the gapped region (in the present terminology, plateau region) [Solyom-Ziman,Inoue-Nomura]. The fixed point of the PRG equation $L\Delta(L) = (L+2)\Delta(L+2)$ is $\Delta(L) \sim 1/L$. For the second order phase transition, the relation $\Delta(L) \sim 1/L$ holds only at the phase transition point, which explains the high reliability of the PRG. On the other hand, for the BKT transition, $\Delta(L) \sim 1/L$ holds in the whole gapless region. Thus the fixed point of the PRG equation is controlled by the higher order terms such as $1/L^2$. This is the reason of the overestimation of the gapped region in the PRG analysis for the BKT transition. Thus, we think, our phase diagram of Fig.4 shows the lower limit of the plateau region.

One of the most reliable methods for analyzing the numerically diagonalization data is the level spectroscopy (LS) method. The LS method for the BKT transition of the present type (namely, without the double degeneracy) was established by Nomura and Kitazawa [Nomura-Kitazawa]. In fact, for instance, Okamoto and Kitazawa successfully obtained the phase diagram at $1/3$ of the saturation magnetization of the $S = 1/2$ ferromagnetic-ferromagnetic-antiferromagnetic (FFA) trimerized chain [Okamoto-Kitazawa, Kitazawa-Okamoto-1] and $S = 3/2$ chain with the *XXZ* and on-site anisotropies [Kitazawa-Okamoto-2]. However, the present situation is somewhat different from those of above works. In cases of the $S = 1/2$ FFA chain and $S = 3/2$ chain, the lowest states of the unit cell in the subspace of $S_{tot}^z = 1/2$ are unique. On the other hand, those of the present model is doubly degenerate,

$$|L\rangle = |\uparrow\uparrow\downarrow\rangle + e^{2\pi i/3}|\uparrow\downarrow\uparrow\rangle + e^{-2\pi i/3}|\downarrow\uparrow\uparrow\rangle \quad (5)$$
$$|R\rangle = |\uparrow\uparrow\downarrow\rangle + e^{-2\pi i/3}|\uparrow\downarrow\uparrow\rangle + e^{2\pi i/3}|\downarrow\uparrow\uparrow\rangle \quad (6)$$

due to the chirality of the equilateral triangle. Thus, it is not obvious whether the LS method by Nomura and Kitazawa can be applied to the present case or not. The development of the LS method for the present case is strongly desired.

In summary, we have shown that the magnetization plateau of the spin-$1/2$ twisted three-leg spin tube vanishes by the introduction of the easy-plane anisotropy by estimating the lower limit boundary of the plateau region for the first time. More accurate determination of the phase diagram is left for future works.

## Acknowledgments

This work has been partly supported by Grants-in-Aids for Scientific Research (Nos. 16H01080(JPhysics), 16K05418, and 16K05419) from the Ministry of Education, Culture, Sports, Science and Technology of Japan (MEXT), and Hyogo Science and Technology Association. We also thank the Supercomputer Center, Institute for Solid State Physics, University of Tokyo.